\documentclass[a4paper,fleqn,usenatbib]{mnras}

\usepackage{newtxtext,newtxmath}

\usepackage[T1]{fontenc}
\usepackage{ae,aecompl}

\usepackage{graphicx}	
\usepackage{amsmath}	

\graphicspath{{Figures/}}

\title[Novel bivariate autoregressive model for predicting and forecasting irregularly observed time series]{Novel bivariate autoregressive model for predicting and forecasting irregularly observed time series}

\author[Elorrieta et al.]{
Felipe Elorrieta,$^{1,3,5}$\thanks{E-mail: felipe.elorrieta@usach.cl}
Susana Eyheramendy,$^{2,3}$
Wilfredo Palma,$^{3}$
Cesar Ojeda$^{4}$
\\
$^{1}$Departmento de Matem\'aticas, Facultad de
  Ciencia, Universidad de Santiago de Chile, Av. Libertador Bernardo 
  O'Higgins 3663. Estacion Central, Santiago, Chile\\
$^{2}$ Faculty of Engineering and Sciences, Universidad Adolfo Iba\~nez, Diagonal Las Torres 2700, Pe\~nalol\'en\\
$^{3}$ Millennium Institute of Astrophysics , Nuncio Monsenor Sotero Sanz 100, Of. 104, Providencia, Santiago, Chile \\
$^{4}$ Escuela de Estad\'istica, Universidad del Valle, Cali, Colombia\\
$^{5}$ Center for Interdisciplinary Research in Astrophysics and Space Exploration (CIRAS), Universidad de Santiago de Chile, Av. Libertador Bernardo O'Higgins 3363, \\
Estación Central, Chile.
}

\date{Accepted XXX. Received YYY; in original form ZZZ}

\pubyear{2020}

\begin{document}
\label{firstpage}
\pagerange{\pageref{firstpage}--\pageref{lastpage}}
\maketitle

\begin{abstract}

In several disciplines it is common to find time series measured at irregular observational times. In particular, in astronomy there are a large number of surveys that gather information over irregular time gaps and in more than one passband. Some examples are Pan-STARRS, ZTF and also the LSST. However, current commonly used time series models that estimate the time dependency in astronomical light curves consider the information of each band separately (e.g, CIAR, IAR and CARMA models) disregarding the dependency that might exist between different passbands. In this paper we propose a novel bivariate model for irregularly sampled time series, called the bivariate irregular autoregressive (BIAR) model. The BIAR model assumes an autoregressive structure on each time series, it is stationary, and it allows to estimate the autocorrelation, the cross-correlation and the contemporary correlation between two unequally spaced time series. We implemented the BIAR model on  light curves, in the g and r bands, obtained from the ZTF alerts processed by the ALeRCE broker. We show that if the light curves of the two bands are highly correlated, the model has more accurate forecast and prediction using the bivariate model than a similar method that uses only univariate information. Further, the estimated parameters of the BIAR are useful to characterize Long Period Variable Stars and to distinguish between classes of stochastic objects, providing promising features that can be used for classification purposes.

\end{abstract}




\section{Introduction}
\label{sec:intro}

In has become increasingly important to have appropriate models for irregularly observed time series. In particular, in astronomy there are several current and future surveys that are generating light curves from astronomical objects  over a couple of years to a decade. These light curves are time series of the flux measured at different time gaps. Usually this type of data is fitted using models that consider time as a  continuous variable such as the popular Gaussian processes or the CARMA  model \cite{Kelly_2014}.  We have followed an alternative approach by developing time series models for irregularly observed measurements where time is considered as a discrete as opposed to a continuous variable: the IAR model \citep{Eyheramendy_2018} and the CIAR model \citep{Elorrieta_2019}. These models can be compared to the CAR(1) model, i.e. they all fit an autoregressive model of order one. The motivation for developing irregularly observed time series models where time is discrete in the field of astronomy, comes from the fact that in this discipline observations can be days, months or even years apart, and therefore fitting a differential equation model is not the most natural type of model.\\

The greatest progress in estimating time dependency has been made using univariate models, i.e. a single time series. This can be seen as a limitation, since many astronomical surveys are currently producing more than one time series for the same object. Among them, one of the surveys expected to have the greatest impact in the field is the Legacy Survey of Space and Time (LSST, \cite{Ivezic_2019}). The LSST will take observations from the southern sky using six optical filters $(ugrizy)$ generating six light curves for the same object. Another survey is the Zwicky Transient Facility (ZTF, \cite{Bellm_2018}) which is currently sending alerts in $g$ and $r$ filters in real time to the community brokers (e.g. ALeRCE \cite{Forster_2020}, Antares, Lasair, MARS). Similarly, there is a set of other surveys in operation, such as the Sloan Digital Sky Survey (SDSS;  \cite{York_2000}) or the Panoramic Survey Telescope and Rapid Response System (Pan-STARRS \cite{Kaiser_2002}), which use photometric system composed of five optical filters.\\ 

Most of the developments that incorporate information from several bands aim to estimate periods of variable stars. For example, \citet{VanderPlas_2015} proposes an extension of the Lomb-Scargle periodogram for multiband light curves. Other examples are given by \citet{Mondrik_2015} which extend the multiharmonic analysis of variance (AOVMHW) method to estimate periods using the information of multiple bands, \citet{Saha_2017} propose an hybrid periodogram which combines the Lomb-Scargle periodogram and the Lafler-Kinman approach and \citet{Huijse_2018} use the quadratic mutual information from several bands to estimate periods. Few methods address the problem of modelling the time dependence between the light curves of different filters from an astronomical object. 
Recently, \citet{Hu_2020} propose a multivariate damped random walk process to model the stochastic variability of light curves from multiple bands. \citet{Mudelsee_2014} follows a similar approach to define an autoregressive bivariate process. In psychological data,
\citet{Haan-Rietdijk_2017}  proposes a multivariate generalization of the CAR(1) model, called the Continuous Vector Autoregressive (CVAR) model.\\ 

In this work, we propose a new model that can fit jointly two irregularly observed time series enabling the estimation of their bivariate structure. We name the model Bivariate Irregular Autoregressive (BIAR). The model depends on three parameters: the autocorrelation, cross-correlation and contemporary correlation.  The formulation of the BIAR model is an extension from the CIAR model (\citet{Elorrieta_2019}). In comparison with the methods that model multiband data (mentioned in the paragraph above), the main advantage of the proposed model is that it allows to estimate both positive and negative values of the autocorrelation parameter of a bivariate irregularly sampled time series. The CAR(1) model and its multivariate versions can only estimate positive autocorrelation. Time series with negative autocorrelation can be found in astronomical data. For example, in \cite{Elorrieta_2019}) we showed several examples of periodic variable star light curves whose residuals, after an harmonic model fit, were negatively correlated. \\

The BIAR model is able to fit two time series measured at the same observational times. In order to satisfy this assumption, we propose two imputation methods to fill the missing values in a bivariate time series. First, we propose a smoothing method based on the BIAR model. Alternatively, we also propose to use an interpolation predictor based on the IAR model (\citet{Eyheramendy_2018}). Based on these smoothing methods, we propose a two-stage procedure to estimate the cross-dependency structure between two unequally spaced time series. In the first stage we match the observation times of the two time series, where the missing values are imputed using the smoothing procedures mentioned above. In the second stage, we estimate the time dependence between the time series using the BIAR model.\\ 

An analysis of the light curves observed in multiple bands simultaneously can be useful to characterize some astronomical objects. A correlated multi-band variability is expected between light curves for long period variable stars (e.g. MIRA, Semi-Regular or Irregular) or stochastic astronomical objects, such as Active Galactic Nuclei (AGN), Quasars or Blazars). For stochastic objects, some discussion about the behavior between the light curves of multiple bands can be found in \cite{Vaughan_2003}, \cite{Chatterjee_2008}, \cite{Kaliswal_2017}, etc. \\

The structure of this paper is as follows. In \S~\ref{sec:methods} we present the Complex irregular autoregressive (CIAR) model  and the Bivariate irregular autoregressive (BIAR) model. For the BIAR model the maximum likelihood estimation via a Kalman Filter is described. In \S~\ref{sec:simulation} we study the accuracy of the estimated parameters of the BIAR model using Monte Carlo simulations. In \S~\ref{sec:Smoothing} we introduce two new smoothing methods and assess their performance on simulated data. In addition, we also propose a procedure to equalize the observation times of the light curves measured in two different bands for the same object. In \S~\ref{sec:Application} we illustrate two applications of the BIAR model using light curves from the ZTF survey. We end this paper with a discussion in \S~\ref{sec:discussion}. \\ 
 
\section{Methods}
\label{sec:methods}

\subsection{Complex Irregular Autoregressive Model}

The Complex Irregular Autoregressive (CIAR) model for unequally spaced time series is an autoregressive model that can estimate both positive and negative autocorrelations. This is achieved by representing a time series as a complex valued sequence with a latent imaginary part. Let $x_{t_j}$ be complex valued time series, the the CIAR model is  defined as,

\begin{equation}  \label{CIAR} 
x_{t_j}=\phi^{t_j-t_{j-1}} \, x_{t_{j-1}} + \sigma_{t_j} \varepsilon_{t_j}, 
\end{equation} 

\noindent where $x_{t_j}=y_{t_j}+ i z_{t_j}$. Note that $y_{t_j}$ and $z_{t_j}$ represent the real and the imaginary part of the complex valued sequence $x_{t_j}$ respectively. In addition, $\phi = \phi^R + i \phi^I$ is the complex autocorrelation coefficient, where $\phi^R$ is the real part of the coefficient and $\phi^I$ is the imaginary part, the standard deviation of the model is given by   $\sigma_{t_j} = \sigma \, \sqrt{ 1-\left|\phi^{t_j-t_{j-1}}\right|^2}$ and $\varepsilon_{t_j} = \varepsilon_{t_j}^R + i \varepsilon_{t_j}^I$ is the complex white noise, with $\varepsilon_{t_j}^R$ and $\varepsilon_{t_j}^I$ assumed to be independent with zero mean and variances $\sigma^2_R$ and $\sigma^2_I$ respectively (in general, $\sigma^2_R = \sigma^2_I =1$ is assumed). Note that the CIAR model can also be written as,

\begin{equation}  \label{CIAReq} 
y_{t_j}+ i z_{t_j}= (\phi^R + i \phi^I)^{t_j-t_{j-1}} \, (y_{t_{j-1}} + i z_{t_{j-1}}) + \sigma_{t_j}(\varepsilon_{t_j}^R + i \varepsilon_{t_j}^I), 
\end{equation} 

In addition, the state-space representation of the CIAR model is given by,

\begin{equation}  
\left(\begin{array}{c} y_{t_j} \\  z_{t_j} \end{array} \right)= \left(\begin{array}{cc} \alpha_{t_j}^{R}  & -\alpha_{t_j}^{I} \\ \alpha_{t_j}^{I} & \alpha_{t_j}^{R} \end{array} \right)\left(\begin{array}{c} y_{t_{j-1}} \\  z_{t_{j-1}} \end{array} \right) + \sigma_{t_j} \left(\begin{array}{c} \varepsilon_{t_j}^R \\ \varepsilon_{t_j}^I \end{array} \right)
\label{CIARSS}
\end{equation}
\begin{equation}  \label{CIARSS2}
y_{t_j}  = \left(\begin{array}{cc} 1 & 0 \end{array} \right) \left(\begin{array}{c} y_{t_j}  \\  z_{t_j}  \end{array} \right)
\end{equation}

\noindent where $\alpha_{t_j}^{R}$ and $\alpha_{t_j}^{I}$ are the reparametrized autocorrelation coefficient of the model, given by $\alpha_{t_j}^{R} = |\phi|^{\delta_{j}} \cos(\delta_{j} \psi)$ and $\alpha_{t_j}^{I} = |\phi|^{\delta_{j}} \sin(\delta_{j} \psi)$, $\delta_{j} = t_j-t_{j-1}$ and $\psi = \arccos\left( \frac{\phi^{R}}{|\phi|}\right)$. Note that the observation matrix $G$ of the state-space representation \eqref{CIARSS} - \eqref{CIARSS2} is defined by $G=\left(\begin{array}{cc} 1 & 0 \end{array} \right)$. This construction of $G$ implies that only  $y_{t_j}$ is observed, i.e. the real part of the process $x_{t_j}$. Consequently, $z_{t_j}$ is a latent process.\\ 

The CIAR model is weakly stationary and its state-space representation is stable under the assumption of $|\phi|=|\phi^R + i \phi^I|=\sqrt{{\phi}^{2R}+ {\phi}^{2I}}<1$.\\

The CIAR model can be extended to include measurement errors of the time series. This is an important issue for the light curves of astronomical objects, as they are commonly measured with instrumental errors. Suppose that $y_{t_j}$ is a CIAR process measured with an error $\omega_{t_j}$ and $y_{t_j}^*$ is the true value of $y_{t_j}$. The measurement error can be incorporated into the model by modifying the equation \eqref{CIARSS2} of the state-space representation of the CIAR model by,

\begin{equation}  \label{CIARSS3}
y_{t_j}  = \left(\begin{array}{cc} 1 & 0 \end{array} \right) \left(\begin{array}{c} y_{t_j}^*  \\  z_{t_j}  \end{array} \right) +  \omega_{t_j}
\end{equation}

\noindent where $\omega_{t_j}$ is the measurement error of $y_{t_j}$ with known variance $\sigma^2_{\omega_y}$.\\

\subsection{Bivariate Irregular Autoregressive Model} 

As mentioned above, the CIAR model assumes that only the real part of the process is observable. Suppose now $X_{t_j}=\left(\begin{array}{c} y_{t_j} \\  z_{t_j}\end{array} \right)$ is a CIAR process with both real and imaginary part observable, we say that $X_{t_j}$ is a bivariate irregular autoregressive process (BIAR), defined by the equations,

\begin{equation}  \label{BIARSS}
\left(\begin{array}{c} y_{t_j}^{*} \\  z_{t_j}^{*} \end{array} \right)= \left(\begin{array}{cc} \alpha_{t_j} & -\beta_{t_j} \\ \beta_{t_j} & \alpha_{t_j} \end{array} \right)\left(\begin{array}{c} y_{t_{j-1}}^{*} \\  z_{t_{j-1}}^{*} \end{array} \right) + \left(\begin{array}{c} \varepsilon_{y_{t_j}} \\ \varepsilon_{z_{t_j}} \end{array} \right)
\end{equation}
\begin{equation}  \label{BIARSS2}
\left(\begin{array}{c} y_{t_j} \\  z_{t_j}\end{array} \right) = \left(\begin{array}{cc} 1 & 0 \\ 0 & 1\end{array} \right) \left(\begin{array}{c} y_{t_j}^{*} \\  z_{t_j}^{*}  \end{array} \right) + \left(\begin{array}{c} \omega_{y_{t_j}} \\  \omega_{z_{t_j}} \end{array} \right)
\end{equation}

\noindent where $y_{t_j}$ and $z_{t_j}$ are the two observable time series measured at times $t_j$ with errors $\omega_{y_{t_j}}$  and $\omega_{z_{t_j}}$ respectively. In addition, the state error vector $\varepsilon_{t_j} = \left(\begin{array}{c} \varepsilon_{y_{t_j}} \\ \varepsilon_{z_{t_j}} \end{array} \right)$ is such that $\varepsilon_{t_j} = L_{t_j} \xi_{t_j}$, where $L_{t_j}$ is the lower triangular matrix obtained from the Cholesky decomposition of the state covariance matrix $Q_{t_j}$. $\xi_{t_j}$ is a bivariate normal random variable with mean $\left(\begin{array}{c}0 \\ 0 \end{array} \right)$ and variance $\Sigma_{\xi} = \left(\begin{array}{cc}1 & \rho_{\xi} \\ \rho_{\xi} & 1 \end{array} \right)$    \\ 

The parameters $\alpha_{t_j}$ and $\beta_{t_j}$ are reparametrizations of the original parameters of the model. These parameters are defined as $\alpha_{t_j} = |\phi|^{\delta_{j}} \cos(\delta_{j} \psi)$ and $\beta_{t_j}= |\phi|^{\delta_{j}} \sin(\delta_{j} \psi)$ respectively, where $\phi = \phi^R + i \phi^I$ is the complex autocorrelation coefficient of $X_{t_j}$, $\delta_{j} = t_j-t_{j-1}$ and $\psi= sgn(\phi^{I})\arccos\left( \frac{\phi^{R}}{|\phi|}\right)$ (See Appendix~\ref{sec:biarder} for a full derivation of this result). In the state equation \eqref{BIARSS}, $\alpha_{t_j}$ can be interpreted as a time variant autocorrelation coefficient of order $\delta_j$ of both time series, while $\beta_{t_j}$ can be interpreted as a time variant crosscorrelation coefficient of order $\delta_j$ between both time series. Note that, the BIAR model assumes the same autocorrelation parameter for both time series. \\

In addition, we define $R_{t_j}$ as the covariance matrix of the observation errors $W_{t_j} = \left(\begin{array}{c} \omega_{y_{t_j}} \\  \omega_{z_{t_j}} \end{array} \right)$ such that $R_{t_j}=\left(\begin{array}{cc} \sigma^2_{\omega y}& 0 \\ 0 &  \sigma^2_{\omega z}\end{array} \right)$ and $\sigma^2_{\omega y}$ and $\sigma^2_{\omega z}$ are the known variances of the measurement errors of $y_{t_j}$ and $z_{t_j}$ respectively. The covariance matrix of the state errors $\varepsilon_{t_j} = \left(\begin{array}{c} \varepsilon_{y_{t_j}} \\ \varepsilon_{z_{t_j}} \end{array} \right)$ is defined by $Q_{t_j} = \Sigma(1-\left|\phi^{t_j-t_{j-1}}\right|^2) $, with $\Sigma$ a scale matrix such that $\Sigma= \left(\begin{array}{cc} \sigma_{y}^2 & 0\\  0 & \sigma^2_z \end{array} \right)$, where $\sigma_{y}$, $\sigma_{z}$ are the standard deviations of the processes $y_{t_j}$ and $z_{t_j}$. Note that the variance matrix of the BIAR process $\Sigma_x$ is given by $\Sigma_x = \Sigma \Sigma_{\xi} + R_{t_j}$. Therefore, the existence of correlation between $y_{t_j}$ and $z_{t_j}$ depends on the parameter $\rho_{\xi}$, the so-called contemporary correlation.\\

\subsection{Statistical Properties of the Model} 

The BIAR model is a weakly stationary process under the assumption that the norm of the parameter $\phi$ is less than one, i.e. $|\phi| = |\phi^R + i \phi^I|< 1$. Consider  $X_{t_j}$ the BIAR process with measurement error component and $X_{t_j}^*$ the BIAR process without measurement error, then the moments of the BIAR model are the following:

\begin{enumerate}[a)] 
\item $\mathbb{E}(X_{t_j}) = \mathbb{E}(X_{t_j}^*) =  \left(\begin{array}{c} 0 \\  0 \end{array} \right)$ 
\item $\mathbb{V}(X_{t_j}^*) = \frac{1}{(1-\left|\phi^{t_j-t_{j-1}}\right|^2)} Q_{t_j} = \Sigma \Sigma_{\xi} $\\ 
\item $\mathbb{V}(X_{t_j}) = \Sigma \Sigma_{\xi} + R_{t_j}$\\ 
\item $\gamma_k =  \overline{\phi^{t_{j+k}-t_j}} \mathbb{V}(X_{t_j})$\\ 
\item $\rho_k =  \overline{\phi^{t_{j+k}-t_j}}$\\ 
\end{enumerate}

\noindent
where $\gamma_k$ and $\rho_k$ are the autocovariance and the autocorrelation functions at lag k of the BIAR model respectively. 

\subsection{Estimation} 
\label{ssec:estimation}

For the BIAR model we use the Kalman recursions to get the maximum likelihood estimator of the parameters. Suppose the following general state-space system,

\begin{equation} \label{SSM} 
X_{t_{j}}^* = F_{t_{j}} X_{t_{j}-1}^*  + V_{t_{j}} 
\end{equation} 
\begin{equation} \label{SSM1} 
X_{t_{j}} = G X_{t_{j}}^* + W_{t_{j}} 
\end{equation} 

\noindent where $F_{t_{j}}$ is the transition matrix and $G$ is the observation matrix of the system. In addition the transition error and the observation error are uncorrelated with zero mean and variances $Q_{t_{j}}$ and $R_{t_{j}}$ respectively.\\  

Given an initial state defined by $X_{t_{0}}^* \sim N(\mu,\Sigma)$, the complete log likelihood function of the state-space representation \eqref{SSM}-\eqref{SSM1} (more details in Shumway and Stoffer (2000)), can be defined by,

\begin{equation} \label{ll} 
\begin{split}
\ell(\phi) &= - \frac{1}{2} \log|\Sigma| - \frac{1}{2} X_{t_{0}}^{*'}\Sigma^{-1}X_{t_{0}}^{*} - \frac{1}{2} \mathop{\sum}\limits_{j=1}^{n}\log|Q_{t_{j}}| - \frac{1}{2} \mathop{\sum}\limits_{j=1}^{n}\log|R_{t_{j}}|\\
&  - \frac{1}{2} \mathop{\sum}\limits_{j=1}^{n}(X_{t_{j}}^* - F_{t_{j}} X_{t_{j}-1}^*)'Q_{t_{j}}^{-1}(X_{t_{j}}^* - F_{t_{j}} X_{t_{j}-1}^*) \\
& - \frac{1}{2} \mathop{\sum}\limits_{j=1}^{n}(X_{t_{j}} - G X_{t_{j}}^*)'R_{t_{j}}^{-1}(X_{t_{j}} - G X_{t_{j}}^*)
\end{split}
\end{equation} 

 The kalman recursions of the system \eqref{SSM}-\eqref{SSM1} are defined by

\begin{equation} 
\centering
\begin{split}
\Lambda_{t_j} &= G \Omega_{t_j} G' + R_{t_j} \\
\Theta_{t_j} &= F_{t_j}\Omega_{t_j} G' \\
\Omega_{t_{j+1}} &= F_{t_j} \Omega_{t_j} F_{t_j}' + Q_{t_j}  - \Theta_{t_j} \Lambda_{t_j}^{-1} \Theta_{t_j}' \\
\nu_{t_j}  &= X_{t_j} - G \hat{X}_{t_j}\\
\hat{X}_{t_{j+1}}  &= F_{t_j} \hat{X}_{t_j} +  \Theta_{t_j} \Lambda_{t_j}^{-1} \nu_{t_j}
\end{split}
\label{eq:Kalman}
\end{equation}

For the BIAR model the initial values are $\hat{X}_{t_1} = \left(\begin{array}{c} 0 \\ 0 \end{array} \right)$ and $\Omega_{t_1}  =  \Sigma_x$. The transition matrix is $F_{t_j} = \left(\begin{array}{cc} \alpha_{t_j}^{R} & -\alpha_{t_j}^{I} \\ \alpha_{t_j}^{I} & \alpha_{t_j}^{R} \end{array} \right)$ and the observation matrix is $G= \left(\begin{array}{cc} 1 & 0 \\ 0 & 1 \end{array} \right)$ the variance of the transition error is $Q_{t_j} = \Sigma(1-\left|\phi^{t_j-t_{j-1}}\right|^2)$ and the variance of the observation error is $R_{t_j}=\left(\begin{array}{cc} \sigma^2_{\omega y}& 0 \\ 0 &  \sigma^2_{\omega z}\end{array} \right)$.

\begin{table*}
\centering 
\caption{\em Maximum likelihood estimation of the parameters ${\phi}^R$ and ${\phi}^I$ of the BIAR model. The values $\hat{\phi}_{C1}$ and $\hat{\phi}_{C2}$ are the estimation obtained by the CIAR model of the autocorrelation of each time series of the bivariate process.\label{tab:sim1}} 
\begin{tabular}{rr|rrr|rrr|rrrr}
\hline 
Case & N & ${\phi}^R$ & $\widehat{\phi}^R$ & SD($\widehat{\phi}^R)$  & ${\phi}^I$ & $\widehat{\phi}^I$ & SD($\widehat{\phi}^I)$ & $\hat{\phi}_{C1}$ & $SD(\hat{\phi}_{C1})$ & $\hat{\phi}_{C2}$ & $SD(\hat{\phi}_{C2})$\\   \hline 
  \hline
1 & 30 & 0.7 &  0.6864 & 0.0622 & 0.6 & 0.5833 & 0.0487 & 0.6816 & 0.1271 & 0.687 & 0.096 \\
2 & 30 & -0.7  & -0.6811 & 0.0512 & -0.6 & -0.5843 & 0.061 & -0.6813 & 0.1031 & -0.6851 & 0.1056 \\
3 & 30 & -0.9 & -0.8744 & 0.0765  & 0.3 & 0.2926 & 0.0576 & -0.8778 & 0.1168 & -0.8808 & 0.1019 \\
4 & 30 & 0.9 & 0.8857 & 0.0374  & -0.3  & -0.2912 & 0.0359 & 0.8872 & 0.0617 & 0.8832 & 0.0867 \\ \hline
1 & 100 & 0.7 & 0.6961 & 0.0208 & 0.6 & 0.5956 & 0.0204 & 0.6968 & 0.0245 & 0.6964 & 0.0253 \\
2 & 100 & -0.7 & -0.6944 & 0.021 & -0.6 & -0.5959 & 0.0212 & -0.6967 & 0.0258 & -0.6977 & 0.0248 \\
3 & 100 & -0.9 & -0.8939 & 0.0158   & 0.3 & 0.2982 & 0.0174 & -0.897 & 0.0178 & -0.8966 & 0.0173 \\
4 & 100 & 0.9 & 0.8956 & 0.0169 & -0.3  & -0.2982 & 0.0168 & 0.8963 & 0.019 & 0.8963 & 0.0187 \\ \hline
1 & 300 & 0.7 & 0.699 & 0.0117 & 0.6 & 0.5986 & 0.0119 & 0.699 & 0.014 & 0.6993 & 0.0139 \\
2 & 300 & -0.7 & -0.6981 & 0.0117 & -0.6 & -0.5985 & 0.0116 & -0.6991 & 0.0131 & -0.6988 & 0.014 \\
3 & 300 & -0.9 & -0.8975 & 0.0086  & 0.3 & 0.2997 & 0.0091 & -0.8987 & 0.0099 & -0.8983 & 0.0095  \\
4 & 300 & 0.9 & 0.8984 & 0.0082 & -0.3  & -0.2993 & 0.0091 & 0.8983 & 0.0095 & 0.8987 & 0.0094   \\ \hline
   \hline
\end{tabular}
\end{table*}

From the kalman recursions parameters we can rewrite the log-likelihood function defined above, as

\begin{equation}
\ell(\phi) \propto  - \frac{1}{2} \mathop{\sum}\limits_{j=1}^n \left( \log (|\Lambda_{t_j}|) + \nu_{t_j}' \Lambda_{t_j}^{-1} \nu_{t_j} \right)
\end{equation}

The maximum likelihood estimators of the parameters $\phi^R$ and $\phi^I$ are obtained by maximizing $\ell(\phi)$. In addition, an estimation of the parameter $\rho_{\xi}$ of the BIAR model can be obtained directly. Given the bivariate innovation matrix of the BIAR process $\nu_{t_j} = (\nu_{t_j}^y ~~ \nu_{t_j}^z)$, where $\nu_{t_j}^y$ and  $\nu_{t_j}^z$ are the innovations of the processes $y_{t_j}$ and $z_{t_j}$ respectively, the parameter $\rho_{\xi}$ can be estimated as,

\begin{equation}
\hat{\rho}_{\xi} = \frac{\nu_{t_j}^{y'} \nu_{t_j}^z}{\sqrt{(\nu_{t_j}^{y'} \nu_{t_j}^y)(\nu_{t_j}^{z'} \nu_{t_j}^z) }}
\end{equation}

The proposed estimation method has complexity of order $\mathcal{O}(n)$. We developed scripts to estimate the parameters of the BIAR model in the statistical languages/softwares R and Python. In Appendix~\ref{sec:compubiar} we include a plot that shows that the algorithm  implemented in R satisfy the theoretical complexity of the proposed estimation method.\\ 



\section{Simulation Results}
\label{sec:simulation}

\subsection{Assessing the Estimation Performance of the BIAR Model} 
\label{ssec:Montecarlo} 

We assess the accuracy of the maximum likelihood estimation via Kalman recursions of the BIAR model parameters ${\phi}^R$ and ${\phi}^I$. We perform Monte Carlo experiments based on 1000 repetitions of each simulation. In the first experiment we generate, in each repetition, bivariate sequences with three different sample size $N=30,100~\text{and}~300$ and four combinations of positive and negative values of the parameters ${\phi}^R$ and ${\phi}^I$. In addition, we assume an identity matrix for the state error variance $\Sigma_{\xi}$. The observational times has been generated using a mixture of two exponential distributions, such that, $f(t|\lambda_1,\lambda_2,\omega_1,\omega_2)=\omega_1g(t|\lambda_1)+\omega_2g(t|\lambda_2)$, where $\lambda_1$ and $\lambda_2$ are the means of each distribution and $\omega_1$ and $\omega_2$ its respective weights. For the experiments below we fix $\lambda_1=15$ and $\lambda_2=2$, $\omega_1=0.15$ and $\omega_2=0.85$.\\  

Table \ref{tab:sim1} shows a summary of the results of the first experiment. We note that the proposed methodology produces an accurate estimation of the parameters of the model. Particularly, for the largest sample size we obtain an estimation bias $<0.0025$ for each combination of values of ${\phi}^R$ and ${\phi}^I$. In Table \ref{tab:sim1}, we also add the univariate estimation of the autocorrelation of each time series, obtained with the CIAR model. Note that, in this case, the CIAR model estimates with great precision the parameter ${\phi}^R$ of the BIAR model, with a little more estimation variance. In addition, the parameter estimated by the CIAR model for both time series allowed us to verify the assumption of equal autocorrelation of the processes.\\

\begin{table*}
\centering 
\caption{\em Maximum likelihood estimation of the parameters ${\phi}^R$ and ${\phi}^I$ of the BIAR model generated with a correlation of $\rho_{\xi}$.The values $\hat{\phi}_{C1}$ and $\hat{\phi}_{C2}$ are the estimation obtained by the CIAR model of the autocorrelation of each time series of the bivariate process. \label{tab:sim2}} 
\begin{tabular}{rrr|rrr|rrr|rr|rr} 
  \hline 
Case & N & $\rho_{\xi}$ & ${\phi}^R$ & $\hat{\phi}^R$ & $SD(\hat{\phi}^R)$ & ${\phi}^I$  & $\hat{\phi}^I$ & $SD(\hat{\phi}^I)$ & $\hat{\phi}_{C1}$ & $SD(\hat{\phi}_{C1})$ & $\hat{\phi}_{C2}$ & $SD(\hat{\phi}_{C2})$\\   \hline  \hline 
1 & 300 & 0.9 & 0.7 & 0.7093 & 0.0090 & 0.6 & 0.6081 & 0.0089 & 0.6596 & 0.0203 & 0.7347 & 0.0115 \\
2 & 300 & 0.9 & -0.7 & -0.7099 & 0.0091 & 0.6 & 0.6084 & 0.0088 & -0.7209 & 0.0128 & -0.6801 & 0.0162 \\
3 & 300 & 0.9 & 0.7 & 0.7093 & 0.0088 & -0.6 & -0.6079 & 0.0089  & 0.7347 & 0.0114 & 0.6594 & 0.0200 \\
4 & 300 & 0.9 & -0.7 & -0.7097 & 0.0092 & -0.6 & -0.6082 & 0.0091 & -0.6799 & 0.0163 & -0.7208 & 0.0128 \\
5 & 300 & 0.9 & 0.9 & 0.9059 & 0.0075 & 0.3 & 0.3024 & 0.0065 & 0.8753 & 0.0143 & 0.9181 & 0.0070 \\
6 & 300 & 0.9 & -0.9 & -0.9057 & 0.0073 & 0.3 & 0.3023 & 0.0068  & -0.9040 & 0.0092 & -0.8937 & 0.0108 \\
7 & 300 & 0.9 & 0.9 & 0.9059 & 0.0075 & -0.3 & -0.3025 & 0.0065 & 0.9182 & 0.0068 & 0.8751 & 0.0147 \\
8 & 300 & 0.9 & -0.9 & -0.9056 & 0.0074 & -0.3 & -0.3023 & 0.0070   & -0.8938 & 0.0109 & -0.9039 & 0.0093 \\
   \hline
\end{tabular}
\end{table*}  

In a second experiment, we add correlation on the state error variance matrix $\Sigma_{\xi}$. Consequently, we generated 1000 repetitions of the BIAR process with contemporary correlation $\rho_{\xi}=0.9$ and sample size $N=300$. Table \ref{tab:sim2} shows a summary of the results obtained. Note that when the correlation structure is added, the accuracy of the maximum likelihood estimate of the model parameters is not affected. However, these correlation structure produces that the CIAR model generates poorer estimates of the univariate autocorrelation of each time series.\\

Finally, in order to assess the accuracy of the estimation procedure  of the state errors $\rho_{\xi}$ (contemporary correlation), we perform a third Monte Carlo experiment. In this experiment, we generate 1000 repetitions of a BIAR process with different value of the parameter $\rho_{\xi}$. In addition, we use two different values of the parameters ${\phi}^R$ and ${\phi}^I$  and we define $N=300$ as the sample size of each generated process. Table \ref{tab:rhoest} shows a summary of the results obtained. Note that, the estimation of the correlation parameter is accurate for each combination of the parameters. In addition, we note a slightly higher variability in the estimated values when the magnitude of the correlation is lower.\\

Note that the BIAR model (equations \eqref{BIARSS} - \eqref{BIARSS2}) assumes that the two time series of the bivariate process are measured at the same times. In astronomical light curves this assumption is almost never satisfied. Therefore, the two light-curves need to be interpolated to obtain a bivariate time series with measurements at the same times. In what follows we present two methods to carry out this interpolation.

\begin{table*}
\centering 
\caption{\em Estimation of the parameters ${\phi}^R$, ${\phi}^I$ and $\rho_{\xi}$ of the BIAR model. \label{tab:rhoest}} 
\begin{tabular}{rr|rrr|rrr|rrr} 
  \hline 
Case & N &${\phi}^R$ & $\hat{\phi}^R$ & $SD(\hat{\phi}^R)$ &${\phi}^I$ & $\hat{\phi}^I$ & $SD(\hat{\phi}^I)$ & $\rho_{\xi}$ & $\hat{\rho_{\xi}}$ & $SD(\hat{\rho_{\xi}})$\\   \hline \hline 
1 & 300 & 0.9 & 0.9059 & 0.0075 & 0.3 & 0.3024 & 0.0065 & 0.9 & 0.8854 & 0.0178 \\
2 & 300 & 0.9 & 0.9007 & 0.0080 & 0.3 & 0.3002 & 0.0081 & 0.5 & 0.4952 & 0.0631 \\
3 & 300 & 0.9 & 0.9007 & 0.0078 & 0.3 & 0.3006 & 0.0080 & -0.5 & -0.4947 & 0.0630 \\
4 & 300 & 0.9 & 0.9062 & 0.0070 & 0.3 & 0.3031 & 0.0063 & -0.9 & -0.8851 & 0.0182 \\
5 & 300 & -0.7 & -0.7097 & 0.0092 & -0.6 & -0.6082 & 0.0091 & 0.9 & 0.8920 & 0.0154 \\
6 & 300 & -0.7 & -0.7014 & 0.0107 & -0.6 & -0.6014 & 0.0111 & 0.5 & 0.4964 & 0.0586 \\
7 & 300 & -0.7 & -0.7012 & 0.0110 & -0.6 & -0.6015 & 0.0109 & -0.5 & -0.4968 & 0.0587 \\
8 & 300 & -0.7 & -0.7093 & 0.0094 & -0.6 & -0.6084 & 0.0088 & -0.9 & -0.8922 & 0.0153 \\
   \hline
   \hline
\end{tabular}
\end{table*}

\section{Smoothing Methods} 
\label{sec:Smoothing} 

 Assume that an astronomical object's light curve measured in two different bands can be modelled by  a bivariate process. Some of these measurements will be present in only one band in several observational times. However, a subset of the measurements at different bands can be very close in time. In this case, we can assume that these measurements are taken at the same times. For a given observation, when it is not possible to match an observation at the other band given a small window size, we say that we are at presence of a missing value in one of the time series. In this latter cases we will need to interpolate an estimated value.\\

\subsection{Matching Procedure} 
\label{sec:Matching}

The matching procedure proposed in this work is based on two step. First,  we paired observations in the two bands that are separated by a time interval of less than a given tolerance. For each pair of observations that meet this condition, the new observational time $t_k$  is defined as the mean of the observational times of the two bands, such that, $t_k = \frac{t_{j_1} + t_{j_2}}{2}$, where $t_{j_i}$ is the time of observation in the $i$-th band.\\

The second step is defined for the set of observations that, at the end of this procedure, do not have a match on the other band. In these cases, we must impute the value on the other time series. We propose two methods to make the imputation of each missing value: i) interpolation based on the IAR model; ii) interpolation based on the BIAR model.\\

\subsection{Smoothing IAR} 
\label{sec:SmoothingIAR} 

 The main idea of this method is to estimate the missing value using the one step interpolation predictor of the IAR model, which can be obtained based on the result of the following lemma.\\

\textbf{Lemma 1:} Suppose an univariate IAR process $y_{t_{j}}$ measured at irregular times $t_j$ with $j=1,\ldots,n$. In addition, suppose that we observe $y_{t_{j-1}}$ and $y_{t_{j+1}}$ and we want to estimate $y_{t_{j}}$. The best linear predictor of $y_{t_{j}}$ based on $y_{t_{j-1}}$ and $y_{t_{j+1}}$ is given by,

\begin{equation}  \label{BLP} 
\hat{y}_{t_{j}}= \alpha y_{t_{j-1}} + \beta y_{t_{j+1}}
\end{equation} 

\noindent where $\alpha = \frac{\phi^{\delta_j} \left(1-\phi^{2 \delta_{j+1}}\right)}{1-\phi^{2 (t_{j+1}-t_{j-1})}}$, $\beta = \phi^{\delta_j}  - \alpha \phi^{t_{j+1}-t_{j-1}}$ and $\delta_j = t_{j}-t_{j-1}$.  See Appendix~\ref{sec:lem1} for a proof of this lemma. Note that the interpolation predictor can be obtained for an observational time $t_j$ such that $t_1 < t_j < t_N$.\\

\subsection{Smoothing BIAR} 
\label{sec:SmoothingBIAR} 

Alternatively, we propose a smoothing method based on the BIAR model. This method aims to estimate a missing value from the likelihood function of the BIAR model (equation \eqref{ll}). In order to get the maximum likelihood estimator of the missing value, we assume that the parameters of the model $\phi^R$ and $\phi^I$ must be known. Thus, the first step of the proposed smoothing method is to estimate the parameters of the BIAR model using the observation from observational times that can be paired. Later, the parameters estimated are used as input value to estimate the missing values.\\

For each estimated missing value, we can compute a confidence interval defined by,

\begin{equation} 
\hat{Y}_{t_{j}} \pm z_{1-\alpha/2} \sqrt{\mathbb{V}(e_{t_{j}})}
\label{eq:confidence}
\end{equation}

\noindent where $\hat{Y}_{t_{j}}$ is the estimation of the missing value at time $t_j$ of the time series, $e_{t_{j}} = Y_{t_{j}} - \hat{Y}_{t_{j}}$ is the estimation error (innovation) with variance $\mathbb{V}(e_{t_{j}}) = \sigma^2(1-\left|\phi^{t_j-t_{j-1}}\right|^2)$ and $z_{1-\alpha/2}$ is the $1-\alpha/2$ quantile of the standard normal distribution.\\

\subsection{Smoothing Simulations}
\label{ssec:smoothsim}

To assess the proposed imputation methods we performed Monte Carlo simulations. In this section we show a summary of the results of these experiments.  In the first experiment, we generate a BIAR process with $n=100$ observations and delete a percentage of them of both time series. The observational times were generated using the mixture of exponential distributions defined in Section \ref{sec:simulation}. These times remains fixed for the $m=100$ repetitions of the BIAR processes performed for each combination of the parameters $\phi^R$, $\phi_I$ and $\rho_{\xi}$.\\
  
The evaluation measure that we use in this experiment is the mean squared error (MSE) between the real and the estimated value. We compute the MSE using the two smoothing methods proposed in this work: the Smoothing IAR ($MSE_{I}$) and the Smoothing BIAR ($MSE_{B}$). In Table \ref{tab:smoothsim} we can see that we obtain small values of the MSE with the BIAR smoothing method proposed in this work. On the other hand, the smoothing method of the IAR model decreases its performance for large values of the cross-correlation parameter $({\phi}^I)$ or in the presence of negative autocorrelation $({\phi}^R)$ in the time-series. Finally, in this experiment, a correlation structure different to zero does not produce a significant variation in the estimated values of the mean squared error. These results are similar when both 5\% and 10\% of the data are removed.\\

\begin{table*}
\centering 
\caption{\em Mean Squared Error of the estimation by maximum likelihood of the missing points on simulated BIAR process. \label{tab:smoothsim}} 
\begin{tabular}{r|rrrrr|rr|rr|} 
  \hline 
Case & N & ${\phi}^R$ & ${\phi}^I$ & $\rho_{\xi}$ & Miss & $MSE_{B}$ & $SD(MSE_{B})$ & $MSE_{I}$ & $SD(MSE_{I})$\\ 
  \hline
1 & 100 & 0.9 & 0.3 & 0 & 5\% & 0.0466 & 0.0238 & 0.0663 & 0.0402 \\
2 & 100 & 0.7 & 0.6 & 0 & 5\% & 0.064 & 0.0345 & 0.2201 & 0.1235 \\
3 & 100 & -0.9 & -0.3 & 0 & 5\% & 0.0598 & 0.0333 & 0.8874 & 0.465 \\
4 & 100 & -0.7 & -0.6 & 0 & 5\% & 0.0822 & 0.0491  & 0.8347 & 0.4495 \\ \hline
5 & 100 & 0.9 & 0.3 & 0.9 & 5\% & 0.0427 & 0.0295 & 0.0621 & 0.0423 \\
6 & 100 & 0.7 & 0.6 & 0.9 & 5\% & 0.0656 & 0.0445 & 0.2402 & 0.1567 \\
7 & 100 & -0.9 & -0.3 & 0.9 & 5\% & 0.0551 & 0.0377& 0.8405 & 0.5283 \\
8 & 100 & -0.7 & -0.6 & 0.9 & 5\% & 0.066 & 0.0381 & 0.8065 & 0.4577 \\ \hline
9 & 100 & 0.9 & 0.3 & 0 & 10\%  & 0.0782 & 0.0347 & 0.1834 & 0.1066 \\
10 & 100 & 0.7 & 0.6 & 0 & 10\% & 0.115 & 0.0475 & 0.4407 & 0.1916 \\
11 & 100 & -0.9 & -0.3 & 0 & 10\% & 0.0932 & 0.0388 & 0.7614 & 0.3187 \\
12 & 100 & -0.7 & -0.6 & 0 & 10\% & 0.1255 & 0.0487 & 0.7248 & 0.3009 \\ \hline
13 & 100 & 0.9 & 0.3 & 0.9 & 10\% & 0.0773 & 0.0451 & 0.1817 & 0.1039 \\
14 & 100 & 0.7 & 0.6 & 0.9 & 10\% & 0.109 & 0.0682 & 0.4001 & 0.2089  \\
15 & 100 & -0.9 & -0.3 & 0.9 & 10\% & 0.0818 & 0.0439  & 0.7476 & 0.292 \\ 
16 & 100 & -0.7 & -0.6 & 0.9 & 10\%  & 0.1259 & 0.0691 & 0.717 & 0.2657 \\ \hline
   \hline
\end{tabular}
\end{table*}

In a second experiment, we delete observations of only one time series of the bivariate process. Consequently, the remaining time series does not have missing values. The main idea of this experiment is that if both series of the bivariate process are correlated, then the time series that is fully sampled should help in the estimation of the missing values of the remaining time series. In this example, we consider two measurements of the mean square error for the BIAR smoothing method. First, we consider the estimation of the smoothed value in only the time series that has the missing value $(MSE_{B1})$. The second method considers the estimation of the value in both time series ($MSE_{B2}$). Note in Table \ref{tab:smoothsim2} that the mean squared error of the BIAR smoothing method reach the minimum value when the bivariate process have a correlation structure and the estimation is performed in a single time series, using the remaining one as support. \\

\begin{table*}
\centering 
\caption{\em Mean Squared Error of the estimation by maximum likelihood of the missing points in one band on simulated BIAR process. \label{tab:smoothsim2}} 
\begin{tabular}{r|rrrrr|rr|rr|rr|} 
  \hline 
Case & N & ${\phi}^R$ & ${\phi}^I$ & $\rho_{\xi}$ & Miss & $MSE_{B1}$ & $SD(MSE_{B1})$ & $MSE_{B2}$ & $SD(MSE_{B2})$ & $MSE_{I}$ & $SD(MSE_{I})$\\ 
  \hline
1 & 100 & 0.9 & 0.3 & 0 & 5\% & 0.0414 & 0.0311 & 0.0412 & 0.0308 & 0.0501 & 0.0398 \\
2 & 100 & 0.7 & 0.6 & 0 & 5\% & 0.077 & 0.0556 & 0.0759 & 0.0537 & 0.227 & 0.2026 \\
3 & 100 & -0.9 & -0.3 & 0 & 5\% & 0.0607 & 0.0513 & 0.0599 & 0.05 & 1.0302 & 0.7378 \\
4 & 100 & -0.7 & -0.6 & 0 & 5\% & 0.0863 & 0.0548 & 0.0865 & 0.0544 & 0.8628 & 0.7236 \\ \hline
5 & 100 & 0.9 & 0.3 & 0.9 & 5\% & 0.0318 & 0.0221 & 0.0447 & 0.0277 & 0.0697 & 0.0491 \\
6 & 100 & 0.7 & 0.6 & 0.9 & 5\%  & 0.0497 & 0.0393 & 0.0705 & 0.0527 & 0.2552 & 0.2222 \\
7 & 100 & -0.9 & -0.3 & 0.9 & 5\%  & 0.0412 & 0.0303 & 0.0484 & 0.0341 & 0.7548 & 0.5681 \\
8 & 100 & -0.7 & -0.6 & 0.9 & 5\%  & 0.04 & 0.0354 & 0.0464 & 0.0392 & 0.6548 & 0.4927 \\ \hline
9 & 100 & 0.9 & 0.3 & 0 & 10\%  & 0.0789 & 0.0416 & 0.0781 & 0.0411 & 0.1748 & 0.1118 \\
10 & 100 & 0.7 & 0.6 & 0 & 10\% & 0.1425 & 0.0915 & 0.1403 & 0.0876 & 0.4742 & 0.3097 \\
11 & 100 & -0.9 & -0.3 & 0 & 10\%  & 0.0877 & 0.0509 & 0.0869 & 0.0499 & 0.7133 & 0.3525 \\
12 & 100 & -0.7 & -0.6 & 0 & 10\% & 0.1351 & 0.0864 & 0.1339 & 0.0849 & 0.7333 & 0.3523 \\ \hline
13 & 100 & 0.9 & 0.3 & 0.9 & 10\% & 0.0649 & 0.0467 & 0.0877 & 0.0623 & 0.1764 & 0.1316 \\
14 & 100 & 0.7 & 0.6 & 0.9 & 10\% & 0.0693 & 0.0432 & 0.0933 & 0.0507 & 0.562 & 0.3888 \\
15 & 100 & -0.9 & -0.3 & 0.9 & 10\% & 0.0582 & 0.0417 & 0.0727 & 0.0488 & 0.7694 & 0.4077 \\
16 & 100 & -0.7 & -0.6 & 0.9 & 10\% & 0.0669 & 0.0467 & 0.0984 & 0.0746 & 0.7344 & 0.4083 \\  \hline
   \hline
\end{tabular}
\end{table*}

\begin{center}
\begin{figure*}
\begin{minipage}{0.49\linewidth}
\includegraphics[width=\textwidth]{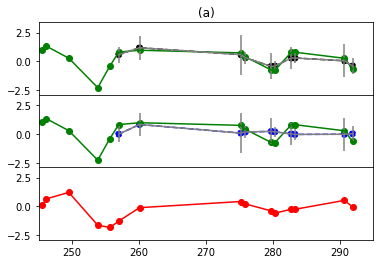}
\end{minipage}
\begin{minipage}{0.49\linewidth}
\includegraphics[width=\textwidth]{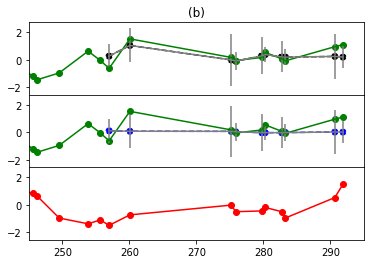}
\end{minipage}
\begin{minipage}{0.49\linewidth}
\includegraphics[width=\textwidth]{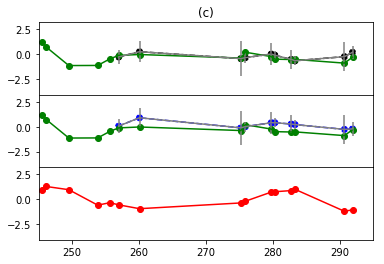}
\end{minipage}
\begin{minipage}{0.49\linewidth}
\includegraphics[width=\textwidth]{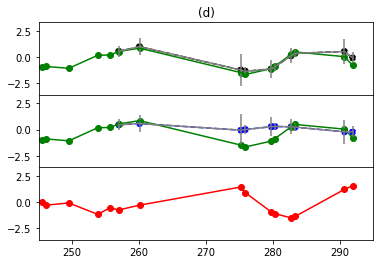}
\end{minipage}
\caption{Forecasting Performance of the BIAR process. In each figure three panels are shown. The first panel compares the BIAR forecast (black dots) and the real values of the time series (green dots). The second panel compares the CIAR forecast (blue dots) and the real values of the same time series. The third panel shows the remaining time series (red dots) of the bivariate process. \label{fig:fore}. Each figure represents a BIAR process simulated with parameters $\phi^R=0.9$ y $\phi^I=0.3$. In addition, the parameter $\rho_{\xi}$ varies between the figures. In Figure a) $\rho_{\xi}$ is defined as 0.9. Figure b) shows a BIAR process with $\rho_{\xi}=0.5$. Figure c) shows a BIAR process with $\rho_{\xi}=0$. Figure d) shows a BIAR process with $\rho_{\xi}=-0.9$}
\end{figure*}
\end{center}

Following the same idea described above, if the light curves of different band are correlated, and one of them does not have recent observations, we can forecast their future values supported by the remaining light curve which has more recent measurements. This can be achieved by using the BIAR smoothing method proposed in this work.  In order to evaluate the impact of using this methodology in the forecasting process we formulated a final experiment. To carry out this experiment, we remove the final 10\% of the measurements of one time series of the bivariate process. Later, we forecast the future values of the incomplete time series using the smoothing method based on the BIAR model. Each forecast is obtained at one step ahead. Then, the predicted values at two or more steps ahead are obtained based on its respective previous estimate. Figure \ref{fig:fore} illustrates the forecast obtained by the BIAR model in different scenarios of correlation between light curves. In order to compare the forecast performance of the BIAR model, we also use the CIAR forecast formula (more details in \cite{Elorrieta_2019}) to obtain the univariate prediction. Note that, both the univariate and the bivariate forecast method have good performance in the absence of contemporary correlation. However, if both time series are correlated, the performance of the univariate forecast gets worse.\\

\section{Application of the BIAR model on Astronomical Data} 
\label{sec:Application} 

In order to illustrate an application of the BIAR model in astronomical time series we use a set of light curves from the Zwicky Transient Facility (ZTF) \citep{Bellm_2018} survey. The dataset with the ZTF alerts that we use in this section was processed by the ALeRCE broker \citep{Forster_2020}.  As mentioned in Section \ref{sec:intro}, this data is appropriate to implement the BIAR model, since the ZTF observes in two bands, the r and the g-band. In particular, we consider a subset of astronomical objects from the ZTF survey which satisfy some conditions. For example, we do not include in our analysis astronomical objects that were detected in a single band or that have no intersection between the observation times of the detections in both bands. In addition, we consider only objects with more than fifty observations in each band. Finally, we consider objects with an univariate estimation of the autocorrelation in each band greater than a tolerance of 0.005. The data set filtered with these conditions has light curves from 9 different classes of astronomical objects. A summary of the data set used in this work by class is shown in Table~\ref{tab:suma}. In addition, we grouped these objects into the following three types: long and short period variable stars (hereinafter just called periodic), and stochastic objects. Note that the dataset is unbalanced with classes with $\approx 1800$ objects such as, the long period variable (LPV) stars, while the stochastic objects total 316 among its five classes. In addition, we also notice a different length of light curves between the class of objects. LPV light curves have on average 100 observations approximately in each band. On the other hand, periodic and stochastic light curves have around 80 observations in each band.\\ 

\begin{table}
\centering
\begin{tabular}{rrrrr}
  \hline
Class & Type & N & $\bar{n}_g$ & $\bar{n}_r$\\
  \hline
  LPV & LPV & 1857 & 98.45 & 118.2 \\ \hline
  AGN & Stochastic & 53 & 71.92 & 66.06 \\
  Blazar & Stochastic & 79 & 81.46 & 85.63 \\
  CV/Nova & Stochastic & 78 & 96.78 & 92.67 \\
  QSO &  Stochastic & 59 & 70.8 & 66.85 \\
  YSO & Stochastic & 47 & 95.04 & 127.96 \\  \hline
& Total Stochastic  & 316 & 83.67 & 86.87 \\ \hline
  Eclipsing & Periodic & 103 & 73.12 & 70.61 \\
  RRL & Periodic & 952 & 91.94 & 77.32 \\
  Periodic-Other & Periodic  & 8 & 65.75 & 64.12 \\ \hline
& Total Periodic  & 1063 & 90.0 & 76.61   \\ \hline
\end{tabular}
\caption{Summary of objects by classes in the dataset. \label{tab:suma}}
\end{table}

Before fitting the BIAR model, we pre-process the data available. The pre-processing considers two stages. First, in order to fit light curves with a stationary behavior, we remove the trend from the time-series. The trend was estimated using the loess method of local regression implemented in the {\em sklearn} package of Python. Later, we implemented the matching procedure explained in Section~\ref{sec:Matching}, between the g-band and r-band light curves for each object. Generally, we found observations that do not have a match in the light curve of the other band. In this case we need to impute these values. To make the imputation of the missing values in the light curves we used the BIAR smoothing method. Consequently, we first estimate the parameters of the BIAR model using only the observations that have a match in the light curves of both bands. We use these estimated parameters to implement the BIAR smoothing. Finally, we re-estimate the BIAR model parameters in the light curves with imputed missing data.\\   

\begin{center}
\begin{figure*}
\begin{minipage}{0.49\linewidth}
\includegraphics[width=\textwidth]{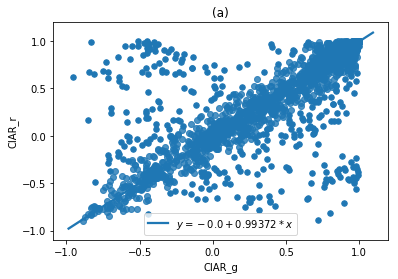}
\end{minipage}
\begin{minipage}{0.49\linewidth}
\includegraphics[width=\textwidth]{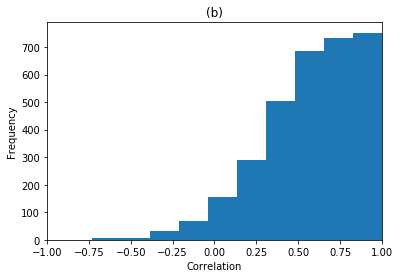}
\end{minipage}
\caption{Figure (a) shows the autocorrelation estimation in the light curves of bands $g$ and $r$. Figure (b) shows the histogram of the Pearson correlation coefficient between the light curves of bands $g$ and $r$. \label{fig:assBIAR}.}
\end{figure*}
\end{center}

After pre-processing the data, we are interested in assessing whether the assumptions of the BIAR model are satisfied. In other words, if both time series have the same autocorrelation structure and if they also have a significant contemporary correlation. To estimate the univariate autocorrelation, we fit the CIAR model to the g and r light curves of each object. On the other hand, to obtain the contemporary correlation, we use the Pearson correlation coefficient. As can be seen in Figure~\ref{fig:assBIAR} (a) there are a direct association between the autocorrelation estimated in the light curves of band g $(CIAR_g)$ and band r $(CIAR_r)$. To determine the degree of association between the estimated autocorrelations in each band, we fit a linear regression to the points near to the diagonal of Figure~\ref{fig:assBIAR} (a).These points correspond to 90.7\% of the objects analyzed. From the fitted model we obtained a slope of 0.99 and an intercept of 0. Figure~\ref{fig:assBIAR} (b) shows that most of the analyzed objects have a significant contemporary correlation between its light curves of both bands. Also, in most of these cases the correlation is positive and close to one. \\

\subsection{Parameters Estimation from the Bivariate Irregular Autoregressive Model} 
\label{sec:Features} 

\begin{center}
\begin{figure*}
\begin{minipage}{0.32\linewidth}
\includegraphics[width=\textwidth]{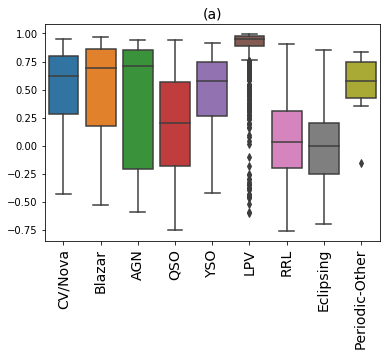}
\end{minipage}
\begin{minipage}{0.32\linewidth}
\includegraphics[width=\textwidth]{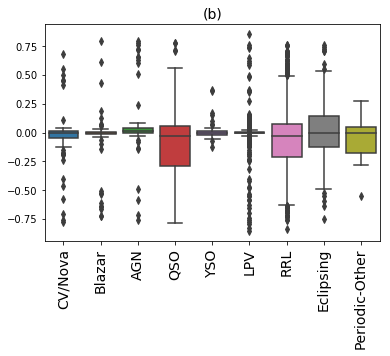}
\end{minipage}
\begin{minipage}{0.32\linewidth}
\includegraphics[width=\textwidth]{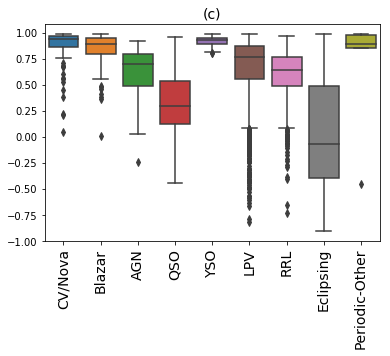}
\end{minipage}
\caption{Boxplot of estimated parameters of the BIAR model by class.Figure (a) shows the boxplot of the estimated values of the parameter $\phi^R$ by class. Figure (b) shows the boxplot of the estimated values of the parameter $\phi^I$ by class. Figure (c) shows the boxplot of the estimated values of the parameter $\rho_{\xi}$ by class.  \label{fig:features}.}
\end{figure*}
\end{center}

From the BIAR model we can obtain three parameters that can be used as features for the classification of objects. These parameters are the autocorrelation of each series fitted by the bivariate model $(\phi^R)$, the crosscorrelation between both series $(\phi^I)$, and the contemporary correlation between the bivariate state error $(\rho_{\xi})$. Figure~\ref{fig:features} shows the estimated parameters by class of astronomical objects in the ZTF data used in this work. Figure \ref{fig:features} (a) shows that the highest estimated values of the autocorrelation parameter are taken by the Long Period Variables. On the other hand, the lower values of this parameter are related to the periodic variable stars, such as eclipsing and RR-Lyrae stars. Regarding to the cross correlation parameter, in Figure  \ref{fig:features}-(b) we can see that for several classes, values close to zero are obtained. However, quasars (QSO) and periodic objects are characterized by negative values in the cross correlation of order one. Finally, Figure~\ref{fig:features}-(c) shows that the contemporary correlation parameter allows to characterize the eclipsing variable stars, since for this class of objects, the lowest estimates of this parameter are obtained. In addition, for stochastic objects such as, CV/Nova, Blazar and YSO the highest estimation of this parameter are obtained.

\begin{center}
\begin{figure}
\begin{minipage}{\linewidth}
\includegraphics[width=\textwidth]{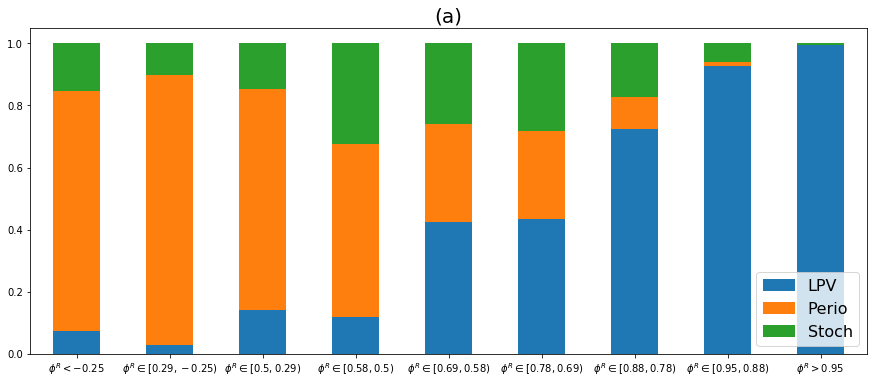}
\end{minipage}
\begin{minipage}{\linewidth}
\includegraphics[width=\textwidth]{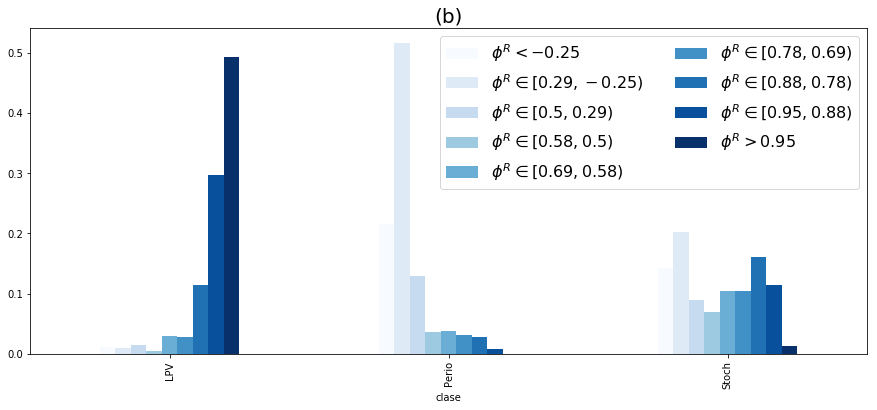}
\end{minipage}
\caption{Barplot of the autocorrelation rank variable $r_{\phi^R}$ by type of object. Figure a) shows the precision plot, in which the x-axis are the values of $r_{\phi^R}$, while the y-axis have the proportion of cases by type of object. Figure b) shows the recall plot, in which the x-axis are the class of objects, while the y-axis have the proportion of cases by value of $r_{\phi^R}$. \label{fig:PrecPhiR}.}
\end{figure}
\end{center}

To obtain an alternative illustration of the results, we grouped the estimated values of the parameter $\phi^R$.The grouping cut-off points were chosen from a decision tree CART algorithm which was implemented using the \textit{tree} function of the \textit{sklearn} package of python \citep{scikit}. We use as explanatory variable the parameter $\phi^R$ and as dependent variable the three type of objects defined in Table \ref{tab:suma}, i.e., LPV, Stochastic and Periodic objects. We set the maximum depth of the tree as 4 and the minimum number of samples required to be at a leaf node as 50. From the computed tree we obtain 9 categories of values of the parameter $\phi^R$. Based on this categories, we define the autocorrelation rank variable $r_{\phi^R}$, which takes the values $1: \phi^R \in [-1,-0.25]$, $2: \phi^R \in (-0.25,0.29]$, $3:  \phi^R \in (0.29,0.5]$, $4:  \phi^R \in (0.5,0.58]$, $5:  \phi^R \in (0.58,0.69]$, $6:  \phi^R \in (0.69,0.78]$, $7:  \phi^R \in (0.78,0.88]$ , $8: \phi^R \in (0.88,0.95]$ and $9:\phi^R \in (0.95,1]$. Figure \ref{fig:PrecPhiR} shows a barplot of the autocorrelation rank variable $r_{\phi^R}$ by type of object. For the ranked variable we show two different plots: one for the precision by class at each threshold, and other for the recall by class. Figure \ref{fig:PrecPhiR} (a) shows that the precision of LPV stars in the rank of estimated values of $\phi^R$ greater than 0.95 is close to one. In other words, if we consider the estimated values of the autocorrelation parameter greater than 0.95 we obtain a practically clean sample of LPV stars. Also note that the amount of LPV stars decreases  together with the value of $\phi^R$ whereas the amount of Periodic stars increases when the value $\phi^R$ decreases.\\

\begin{center}
\begin{figure}
\begin{minipage}{\linewidth}
\includegraphics[width=\textwidth]{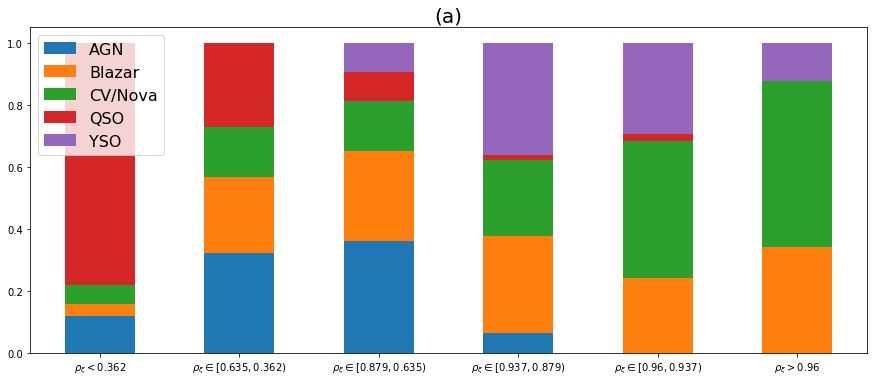}
\end{minipage}
\begin{minipage}{\linewidth}
\includegraphics[width=\textwidth]{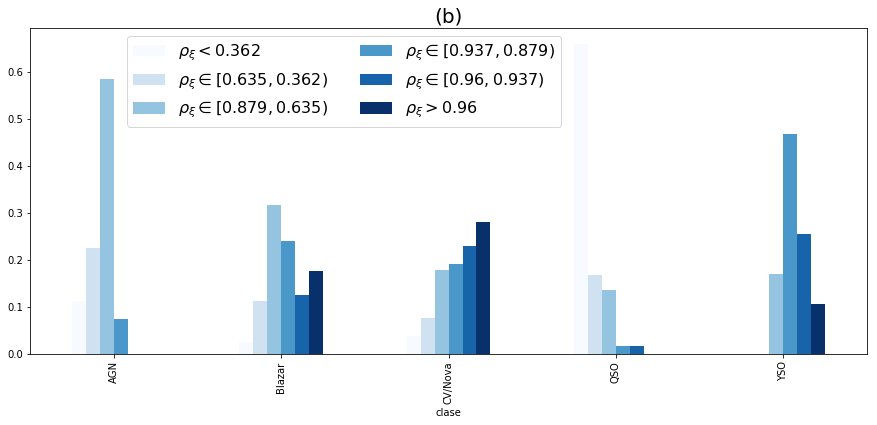}
\end{minipage}
\caption{Barplot of the autocorrelation rank variable $r_{\rho_{\xi}}$ by type of object. Figure a) shows the precision plot, in which the x-axis are the values of $r_{\rho_{\xi}}$, while the y-axis have the proportion of cases by type of object. Figure b) shows the recall plot, in which the x-axis are the class of objects, while the y-axis have the proportion of cases by value of $r_{\rho_{\xi}}$. \label{fig:PrecRho}.}
\end{figure}
\end{center}

Another result that we would like to highlight is the ability of the parameter $\rho_{\xi}$, the contemporary correlation, to separate between the classes of stochastic objects. To illustrate this result, we define the contemporary correlation rank variable $r_{\rho_{\xi}}$. Like the previous example, the categorization of the variable was made from a decision tree. Here we use as an explanatory variable the estimated values of the parameter $\rho_{\xi}$ and as dependent variable we use the five categories of stochastic objects (AGN, Blazars, CV\textbackslash Nova, QSO, YSO). In this case, we set the minimum number of samples required to be at a leaf node as 30 and we kept the maximum depth of the tree as in the previous example. From the decision tree we obtain the following six categories of values, $1:  \rho_{\xi}  \in [-1,0.362]$, $2:  \rho_{\xi} \in (0.362,0.635]$, $3:  \rho_{\xi} \in (0.635,0.879]$, $4: \rho_{\xi}  \in (0.879,0.937]$, $5: \rho_{\xi}  \in (0.937,0.96]$, $6:\rho_{\xi} \in (0.96,1]$. Figure \ref{fig:PrecRho} shows the precision and recall plot of the estimation of the parameter $\rho_{\xi}$ in the light curves of the stochastic objects. Note that for values of $\rho_{\xi}$ higher than 0.96, we find only three of the five stochastic objects analyzed in this work, these are, YSO, CV\textbackslash Nova and Blazars. On the other hand, for the lower values of  $\rho_{\xi}$, we find in a great majority of QSO objects. Figure \ref{fig:PrecRho} b) shows that nearly 85\% of the YSO and CV\textbackslash Nova objects have a $\rho_{\xi}$ estimation greater than 0.88. Finally, the 65\% of the QSO objects have estimation of this parameter lower of 0.36, while the estimates for AGN objects are typically between 0.36 and 0.88.\\

\subsection{Forecasting on ZTF Light Curves} 
\label{sec:ForeZTF} 

A second application of the BIAR model is to forecast future values of light curves. As we discussed in Section \ref{ssec:smoothsim}, a major advantage of the bivariate modeling is that one of the series can support the forecast of the other one. In astronomical data, these series can represent the light curves of the same astronomical object observed in two different bands. The ZTF data that we used in this work consider both g-band and r-band observations.\\

To illustrate the forecast in the ZTF data we selected light curves within the group of stochastic objects. We use the stochastic objects because according to the analysis in Section~\ref{sec:Features}, they have dissimilar values of the parameter $\rho_{\xi}$. For example, the YSO, CV\textbackslash Nova and Blazar objects have higher estimated values of this parameter. On the other hand, the AGN have contemporary correlation estimates around 0.5, while the QSOs have in several cases negative estimates of this parameter. According to the results of the simulations shown previously, when the contemporary correlation parameter is far from zero, the bivariate modeling achieves a better forecast than the one obtained using a univariate approach.\\

\begin{center}
\begin{figure*}
\begin{minipage}{0.48\linewidth}
\centering
\includegraphics[width=8cm,height=6cm]{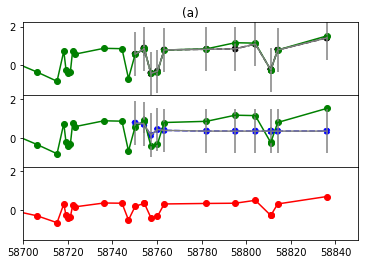}
\end{minipage}
\begin{minipage}{0.48\linewidth}
\centering
\includegraphics[width=8.5cm,height=6cm]{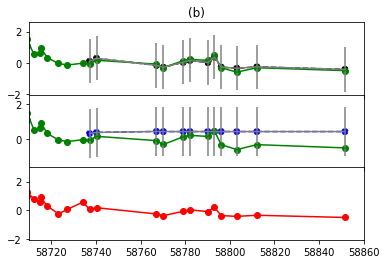}
\end{minipage}
\begin{minipage}{0.48\linewidth}
\includegraphics[width=\textwidth]{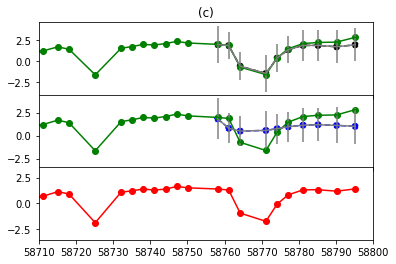}
\end{minipage}
\begin{minipage}{0.48\linewidth}
\includegraphics[width=\textwidth]{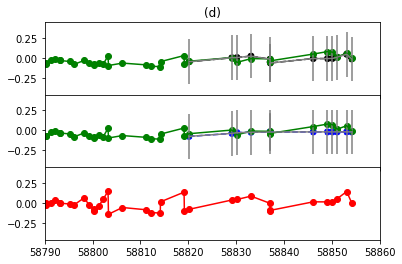}
\end{minipage}
\caption{Forecasting Performance of the BIAR process on the ZTF light curves. In each figure three panels are shown. The first panel compares the BIAR forecast (black dots) and the real values of light curve observed in the g-band (green dots). The second panel compares the CIAR forecast (blue dots) and the real values of the same time series. The third panel shows the light curve of the same object observed in the r-band (red dots) \label{fig:foreZTF}. Figure a) shows the light curves of both bands of a YSO object. Figure b) shows the light curves of both bands of a Blazar object. Figure c) shows the light curves of both bands of a  CV \textbackslash Nova object. Figure d) shows the light curves of both bands of a AGN object.}
\end{figure*}
\end{center}

To assess the forecast performance, we remove the last 10\% of the g-band light curve. Thus, to obtain the forecast value of each observation of the last 10\% of the light curve, we performed an iterative procedure in which each observation is predicted at one step ahead and then replaced by this estimation to forecast the next value. Figure \ref{fig:foreZTF} shows the forecast obtained using the BIAR and the CIAR models in four selected stochastic objects from the ZTF. Figure \ref{fig:foreZTF}-a) shows the forecast in a YSO object from ZTF with an estimated parameter $\rho_{\xi}$ of 0.99. Note that, the forecasted values using the BIAR model are closer to the true value with respect to the forecast obtained using the univariate approach. We obtain similar results in a Blazar light curve with an estimated $\rho_{\xi}$ of 0.93 (Figure \ref{fig:foreZTF}-b)) and a CV \textbackslash Nova light curve with an estimated parameter $\rho_{\xi}$ of 0.99 (Figure \ref{fig:foreZTF}-c)). The most similar performance between the univariate and bivariate approach was obtained in Figure \ref{fig:foreZTF}-d). Here the forecast was performed to an AGN object with a contemporary correlation parameter equal to 0.23. Note that in Figures~\ref{fig:foreZTF}-a), b) and c)  if a pair of light curves have high contemporary correlation, the forecast obtained from the BIAR model is consistently better than that obtained with an univariate model. Furthermore, in Figure \ref{fig:foreZTF}-d) we show in an example with less correlation between the light curves, that the forecast is still satisfactory. The accurate forecasting performance of the BIAR model is due to the fact that the prediction of a future value of one of the time series is based both on the past values of both series of the bivariate process and on the current value of the remaining time series. Consequently, if a light curve has more recent measurements in one passband, these observations can help to better forecast the light curve of the same object in another passband.\\
 
\section{Discussion}
\label{sec:discussion}

We have introduced a novel model to fit bivariate time series that have been observed at irregular discrete times. The model has an autoregressive structure at each univariate series, and it takes into account the correlation between both series. The model contain three parameters. One describes the autocorrelation structure of the two time series, another parameter defines the cross-correlation structure and the third parameter corresponds to the contemporary correlation between both series. The model is stationary, and it allows to do accurate imputation and forecasting.\\

The model assumes that both time series have the same observational times, which is not always attainable. To overcome this limitation we proposed two methods for imputation  in order to end up with two series with measurements at the same times.  One method is based on the IAR model and the other method is based on our proposed BIAR model. Both methods have good performance under some conditions. The IAR model can only estimate positive autocorrelation, and therefore it has good performance when the autocorrelation parameter in both time series is equal and positive. The smoothing method that is based on the BIAR model requires a minimum set of equal observational times in both time series in order to accurately estimate  the parameter of the model. In Section~\ref{sec:simulation} we show that for a bivariate process of 30 observations we can obtain an accurate estimate of the parameters of the BIAR model. When this is achieved, the method gives good performance. We are aware that there are many other methods that can be implemented to impute missing observations that could be implemented as well in this context (e.g. \cite{Crassidis_2011}). Comparison with those methods is out of the scope of this work. \\ 

A key assumption of our model is that both time series have equal autocorrelation. Even though that might appear as a stringent constraint, it is quite natural in astronomical time series when the luminosity is measured on two different filters as we have shown here. We have observed that a large majority of the time series that we analyzed from ZTF, satisfy this condition.\\

We implemented the BIAR model on light curves observed from ZTF in two different passbands for the same astronomical object. We show that this model allows us to detect correlation structure between the g-band and r-band light curves for most of the objects analyzed. Particularly, we show that the contemporary correlation parameter allows us to distinguish between the classes of stochastic objects, such that Blazar, YSO and CV \textbackslash Nova have large values of this parameter, while AGN and QSO are characterized by having smaller correlations between their light curves observed in different bands. On the other hand, we show that the autocorrelation parameter is useful to distinguish the long period variable stars. The cross correlation parameter take negative values for the group of short period variable stars and for the QSO objects. Consequently, the parameters estimated from the BIAR model can be used as features for automated classification of light curves. Classifiers implemented in previous studies have generally considered features that come from harmonic models or univariate autoregressive models (see for example, \citet{Debosscher_etal07}, \citet{Richards_etal11}, \citet{Nun_2015}, \citet{Elorrieta_etal16}, \citet{Sanchez_Saez_2021}).\\

In this work we also show that the proposed model is useful for smoothing and forecasting. In our approach the time series that has the most recent observations helps to the estimation of the missing points in the remaining series. Consequently, if two time series are highly correlated we show that performance in both smoothing and forecasting is better than the univariate approach. We illustrate this result by means of both a simulated BIAR process and light curves from ZTF.\\

With this model we have extended time series models to the possibility of fitting bivariate time series observed at irregular discrete times. The alternative multivariate model for unequally spaced time series that can be found in the literature is the multivariate damped random walk (or multivariate CAR(1)) model (\citet{Mudelsee_2014}, \citet{Haan-Rietdijk_2017}, \citet{Hu_2020}). The BIAR model that we propose differs from this model in that, our model  is formulated in a discrete-time representation and can estimate negative values of the autocorrelation of order one for each time series of the bivariate process.  Furthermore, our model estimates only three parameters, while the bivariate damped random walk models available in the literature estimates five. We have shown how our model is specially suitable for fitting light curves from astronomical objects observed at two passbands, but the model can be used in other disciplines as well.\\

The codes to implement the proposed model are available in PYTHON and R in the following github repository: https://github.com/felipeelorrieta

\section*{Data Availability}

The data that was used in this article consists of light curves from ZTF alerts processed by the Alerce broker. These are  available from the Alerce website \texttt{alerce.science}.

\section*{Acknowledgements}

The authors acknowledge support from the ANID – Millennium Science Initiative Program – ICN12\_009 awarded to the Millennium Institute of Astrophysics MAS (FE, SE, WP).FE acknowledges support from the National Agency for Research and Development (ANID) grant Fondecyt Iniciacion \#11200590.

\bibliographystyle{mnras}
\bibstyle{mnras}
\bibliography{MNRASBIAR.bib}

\appendix

\section{Reparametrization of the BIAR Model}
\label{sec:biarder}

Consider the BIAR model,
\begin{equation}  \label{CIAReq2} 
y_{t_j}+ i z_{t_j}= (\phi^R + i \phi^I)^{t_j-t_{j-1}} \, (y_{t_{j-1}} + i z_{t_{j-1}}) + \sigma_{t_j}(\varepsilon_{t_j}^R + i \varepsilon_{t_j}^I). 
\end{equation} 
We will focus on the term $(\phi^R + i \phi^I)^{t_j-t_{j-1}}$, then
\begin{eqnarray*}
(\phi^R + i \phi^I)^{t_j-t_{j-1}}= (\phi^R + i \phi^I)^{\delta_j} = |\phi|^{\delta_j} \left(\frac{\phi^R + i \phi^I}{|\phi|} \right)^{\delta_j}, 
\end{eqnarray*}
where $\phi=\phi^R + i \phi^I$. Using the polar representation of the complex number, we have 
$$\frac{\phi^R}{|\phi|} + i \frac{\phi^I}{|\phi|} = \cos(\psi) +  i \sin (\psi),$$
where, 
$$\psi = \begin{cases} \arccos\left(\frac{\phi^R}{|\phi|} \right), & if  \phi^I \geq 0 \\ 
- \arccos\left(\frac{\phi^R}{|\phi|} \right), & if  \phi^I < 0 \\ 
\end{cases}$$
Now, using De Moivre's formula, we have
\begin{eqnarray*}
|\phi|^{\delta_{j}} (\cos(\psi) + i \sin(\psi))^{\delta_{j}} &=& |\phi|^{\delta_{j}} (\cos({\delta_{j}} \psi) + i \sin({\delta_{j}} \psi))\\
&=& |\phi|^{\delta_{j}} \cos({\delta_{j}} \psi) + i  |\phi|^{\delta_{j}} \sin({\delta_{j}} \psi) \\
&=& \alpha_{t_j}^{R} + i \alpha_{t_j}^{I}.\\
\end{eqnarray*}
Finally, the BIAR model can be represented by the expression,
\begin{eqnarray*}
y_{t_j}+ i z_{t_j}= (\alpha_{t_j}^{R} + i \alpha_{t_j}^{I}) \, (y_{t_{j-1}} + i z_{t_{j-1}}) + \sigma_{t_j}(\varepsilon_{t_j}^R + i \varepsilon_{t_j}^I). \, \Box\\
\end{eqnarray*}

\section{Proof of Lemma 1}
\label{sec:lem1}

A linear predictor of $y_{t_{j}}$ based on $y_{t_{j-1}}$ and $y_{t_{j+1}}$ is given by,
$$\hat{y}_{t_{j}}= \alpha y_{t_{j-1}} + \beta y_{t_{j+1}}.$$
By the projection theorem, we know that,
\begin{eqnarray} 
\mathbb{E}((y_{t_{j}} - \hat{y}_{t_{j}})y_{t_{j-1}}) &=&  0\\ 
\mathbb{E}((y_{t_{j}} - \hat{y}_{t_{j}})y_{t_{j+1}}) &=&  0, 
\end{eqnarray} 
or equivalently,
\begin{eqnarray*} 
\mathbb{E}(y_{t_{j}} y_{t_{j-1}}) &=&  \mathbb{E}(\hat{y}_{t_{j}}y_{t_{j-1}})\\ 
\mathbb{E}(y_{t_{j}} y_{t_{j+1}}) &=&  \mathbb{E}(\hat{y}_{t_{j}}y_{t_{j+1}}).
\end{eqnarray*} 
Note that $\hat{y}_{t_{j}} = \alpha y_{t_{j}-1} + \beta y_{t_{j}+1}$, then,
\begin{eqnarray*} 
\mathbb{E}(y_{t_{j}} y_{t_{j}-1}) &=&  \mathbb{E}((\alpha y_{t_{j-1}} + \beta y_{t_{j}+1})y_{t_{j}-1})\\ 
\mathbb{E}(y_{t_{j}} y_{t_{j}+1}) &=&  \mathbb{E}((\alpha y_{t_{j-1}} + \beta y_{t_{j}+1})y_{t_{j}+1}).
\end{eqnarray*} 
As $\mathbb{E}(y_{t_{j}}) = 0$, then the autocovariance function can be defined by $\gamma_y(t_{j}-t_{j-1}) = \mathbb{E}(y_{t_{j}} y_{t_{j-1}})$. Therefore, 
\begin{eqnarray*} 
\gamma_y(t_{j}-t_{j-1}) &=&  \alpha \gamma_y(0) + \beta\gamma_y(t_{j+1}-t_{j-1})\\ 
\gamma_y(t_{j+1}-t_{j}) &=&  \alpha \gamma_y(t_{j+1}-t_{j-1}) + \beta \gamma_y(0).
\end{eqnarray*} 
The above equations can be rewritten using the autocorrelation function $\rho_y$ as,
\begin{eqnarray*} 
\rho_y(t_{j}-t_{j-1}) &=&  \alpha + \beta\rho_y(t_{j+1}-t_{j-1})\\ 
\rho_y(t_{j+1}-t_{j}) &=&  \alpha \rho_y(t_{j+1}-t_{j-1}) + \beta.
\end{eqnarray*} 
Note that $\rho_y(t_{j}-t_{j-1})= \phi^{(t_{j}-t_{j-1})}$ then,
\begin{equation}\label{eqa1}
\phi^{(t_{j}-t_{j-1})}  =  \alpha + \beta\phi^{(t_{j+1}-t_{j-1})},
\end{equation} 
\begin{equation}\label{eqa2} 
\phi^{(t_{j+1}-t_{j})} =  \alpha \phi^{(t_{j+1}-t_{j-1})} + \beta.
\end{equation} 
By applying \eqref{eqa1} -  $\phi^{(t_{j+1}-t_{j-1})}$\eqref{eqa2} we obtain,
$$\alpha = \frac{\phi^{(t_{j}-t_{j-1})} \left(1-\phi^{2 (t_{j+1}-t_{j})}\right)}{1-\phi^{2 (t_{k+1}-t_{j-1})}},$$
and, finally from \eqref{eqa2} we get the result,
$$\beta = \phi^{(t_{j+1}-t_{j-1})}  - \alpha \phi^{t_{j+1}-t_{j-1}} ~~~ \Box$$\\ 

\section{Computation Time of the BIAR Model}
\label{sec:compubiar}

\begin{figure}
\includegraphics[width=\linewidth]{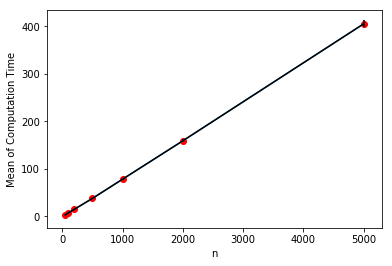}
\caption{Scatter plot between the sample size of the BIAR process and the mean of the computational times of our proposed estimation method. Error bars were calculated as 1.96 times the standard deviation of each mean value. \label{fig:compbiar}}
\end{figure}

In this section, we perform a simulation experiment to analyze how the estimation method that we propose for the BIAR model scales with the data size. In this experiment, we generate 1000 repetitions of the BIAR model. For each repetition calculate the computation time of our implementation in the software R of the proposed estimation method
of the BIAR model based on the Kalman recursions described in the Section~\ref{ssec:estimation}.
In Figure~\ref{fig:compbiar} we show the mean of the computational times calculated for each repetition for different values of the data size. Note that the computation time of our implementation increase linearly with the sample size as expected due to the theoretical complexity of the Kalman recursion.\\

\bsp	
\label{lastpage}
\end{document}